# Unraveling Magneto-Phononic Coupling and Photoinduced Magnetic Control in Antiferromagnetic Kondo Semimetal CeBi

Xu-Chen Nie,[1] Chen Zhang,[2] Qi-Yi Wu,[2] Hao Liu,[2] Jie Pang,[3,4] You-Guo Shi,[3,4] Zhi-An Xu,[5] Yan-Feng Guo,[5,6] Ya-Hua Yuan,[2] H. Y. Liu,[7] Yu-Xia Duan,[2] and Jian-Qiao Meng[2,*]

[1]*Key Laboratory for Intelligent Nano Materials and Devices of Ministry of Education,*
*State Key Laboratory of Mechanics and Control of Mechanical Structures, and Institute for Frontier Science,*
*Nanjing University of Aeronautics and Astronautics, Nanjing 210016, Jiangshu China*
[2]*School of Physics, Central South University, Changsha 410083, Hunan, China*
[3]*Beijing National Laboratory for Condensed Matter Physics,*
*Institute of Physics, Chinese Academy of Sciences, Beijing 100190, China*
[4]*School of Physical Sciences, University of Chinese Academy of Sciences, Beijing 100049, China*
[5]*State Key Laboratory of Quantum Functional Materials,*
*School of Physical Science and Technology, ShanghaiTech University, Shanghai 201210, China*
[6]*ShanghaiTech Laboratory for Topological Physics, ShanghaiTech University, Shanghai 201210, China*
[7]*Beijing Academy of Quantum Information Sciences, Beijing 100085, China*
(Dated: Saturday 12th July, 2025)

We report an ultrafast optical spectroscopy study on the coherent phonon dynamics in the topological semimetal CeBi and its nonmagnetic isostructural compound LaBi, revealing profound insights into their electronic and magnetic interactions. Both materials exhibit prominent $A_{1g}$ longitudinal optical phonons with characteristic anharmonic temperature dependencies. However, in CeBi, the $A_{1g}$ phonon frequency and amplitude show clear anomalies near its antiferromagnetic (AFM) ordering temperatures ($T_{N1} \simeq 25$ K and $T_{N2} \simeq 12$ K), which unequivocally demonstrate strong magneto-phononic coupling. Crucially, in the AFM state at 4 K, CeBi exhibits a pump fluence threshold of $F_C \approx 44$ $\mu$J/cm$^2$, above which the rate of phonon softening accelerates and the amplitude increases sharply. This unique threshold, absent in paramagnetic CeBi and LaBi, points to a photoinduced, non-thermal quenching of the AFM order. Our findings establish coherent phonons as highly sensitive probes of intertwined orders in heavy fermion systems, highlighting the transformative potential of ultrafast pulses in dynamically controlling magnetic states in correlated electron materials and paving the way for the manipulation of emergent quantum phases.

The emergent properties of quantum materials arise from the intricate interplay among fundamental excitations: electronic, lattice, and magnetic. Within this framework, phonons, as quantized lattice vibrations, play a pivotal role in phenomena from conventional superconductivity [1] to complex many-body effects such as the Kondo mechanism [2, 3]. A critical aspect of this interplay is magneto-phononic coupling—the interaction between phonons and magnetic order [4]. This coupling gives rise to emergent phenomena such as colossal magnetoresistance [4], multiferroicity [5], quantum spin liquids [6], and Skyrmionic spin textures [7] in strongly correlated systems. Beyond their direct influence, phonons also serve as sensitive spectroscopic probes, with changes in their characteristics directly revealing quasiparticle interactions and lattice instabilities [8–10]. This makes them indispensable for deciphering the microscopic mechanisms behind symmetry-breaking transitions and emergent electronic orders in various materials [10–14], including heavy fermion compounds [15–17].

Rare-earth monopnictides, characterized by their robust rocksalt crystal structure, constitute a fascinating family of materials that exhibit a rich array of exotic properties, including extremely large magnetoresistance [18, 19], unconventional superconductivity [20], topological insulator behavior [21], Weyl fermions [22], and heavy fermion characteristics [23]. Among these, CeBi stands as an antiferromagnetic (AFM) Kondo semimetal, distinguished by an exceptionally low carrier density of $\sim$0.03 $e^-$/Ce [24]. This limited carrier concentration significantly suppresses Kondo screening, imbuing the Ce-4$f$ electrons with a dual localized-itinerant character [23]. This duality, in turn, gives rise to complex magnetic structures that are highly sensitive to external parameters, evident in a two-step AFM transition at Néel temperatures $T_{N1} \simeq 25$ K and $T_{N2} \simeq 12$ K [25]. Due to the strong localization of Ce-4$f$ electrons in CeBi, its electronic structure bears a remarkable resemblance to its non-4$f$ isostructural compound, LaBi [26, 27]. This structural and electronic similarity, coupled with the distinct magnetic order of CeBi, makes a comparative study of CeBi and LaBi an ideal approach for determining the respective roles of magnetism and $f$-electron correlations in lattice dynamics.

While conventional tuning parameters such as temperature, doping, pressure, and magnetic fields have long been employed, the recent development of ultrafast optical techniques has revealed new possibilities for dynamically manipulating quantum materials [10]. Intense photoexcitation can transiently reconstruct free-energy landscapes by suppressing order parameters, thereby successfully inducing nonequilibrium phase transitions [28]. Ultrafast optical spectroscopy, with its exquisite temporal resolution, is invaluable for directly detecting and characterizing the coupling between magnetic order and phonon modes [16, 29, 30]. It can capture transient low-energy electronic structures and energy transfer processes that are critical to phase evolution. Furthermore, its ability to provide real-time observation of phonon renormalization and precursor fluctuations at phase boundaries makes it an essential tool for comprehensively characterizing intricate transition pathways in correlated quantum materials.

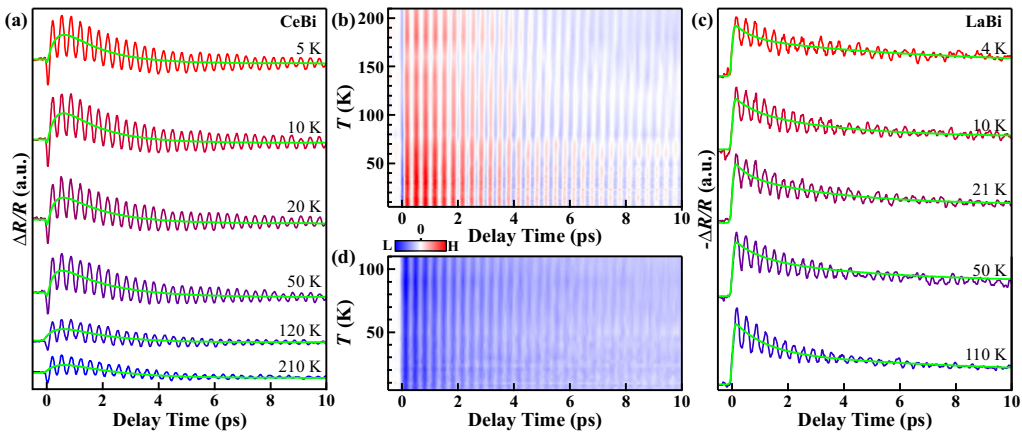

FIG. 1. (a) Differential reflectivity $\Delta R(t)/R$ as a function of delay time for CeBi at various temperatures (pump fluence of $\sim$26.5 $\mu$J/cm$^2$). Green solid curves represent fits to Eq. (1). (b) 2D pseudocolor map illustrating the $\Delta R(t)/R$ as a function of temperature and delay time for CeBi. (c) Differential reflectivity $\Delta R(t)/R$ as a function of delay time for LaBi at various temperatures (pump fluence of $\sim$53 $\mu$J/cm$^2$). Green solid curves represent fits to Eq. (1). (d) 2D pseudocolor map illustrating the $\Delta R(t)/R$ as a function of temperature and delay time for LaBi.

In this Letter, we present ultrafast spectroscopy results obtained from high-quality single crystals of CeBi and its nonmagnetic isostructural compound LaBi. Our measurements reveal the distinct photoexcited carrier dynamics and coherent phonon responses of these materials. We demonstrate that the coherent $A_{1g}$ phonon dynamics in CeBi exhibit significant changes near its magnetic ordering temperatures, $T_{N1}$ and $T_{N2}$, which unequivocally signify strong magneto-phononic coupling. Furthermore, within the low-temperature AFM phase, both the phonon frequency and amplitude exhibit marked anomalies at a critical pump fluence of approximately $F_C \approx$ 44 $\mu$J/cm$^2$. This observation strongly suggests a photoinduced, non-thermal quenching of the AFM order. These findings not only highlight fundamental magneto-phononic coupling mechanisms but also establish new pathways for quantum phase manipulation in correlated materials, demonstrating the power of ultrafast pulses to dynamically control interactions in complex quantum systems.

Ultrafast optical spectroscopy measurements were performed using a Yb-based femtosecond (fs) laser oscillator. This system delivered approximately 35 fs pulses centered at 800 nm ($\sim$1.55 eV) with a repetition rate of 1 MHz. For these measurements, cross-polarized pump and probe beams were focused onto the same spot on a freshly cleaved (001) sample surface at near-normal incidence. The pump beam had an approximate diameter of 170 $\mu$m, while the probe beam was approximately 70 $\mu$m. To minimize environmental interference, all data acquisition was conducted under high vacuum conditions ($10^{-6}$ mbar). Further details regarding the experimental setup are available in our previous publications [31, 32]. High-quality single crystals of CeBi [33] and LaBi [26] were synthesized via solid-state reaction methods utilizing indium as a flux.

As shown in Fig. 1, both materials exhibit distinct transient reflectivity responses following photoexcitation. Specifically, the differential reflectivity, $\Delta R(t)/R$, in both CeBi [Fig. 1(a)] and LaBi [Fig. 1(c)] shows high-frequency oscillations immediately after photoexcitation, which are superimposed on a non-oscillatory decay. In CeBi, these oscillations are associated with an initial drop in reflectivity that then becomes positive after approximately half an oscillation cycle. Conversely, in LaBi, while the oscillations also induce an initial decrease in reflectivity, the signal consistently remains negative throughout the entire observed period. A notable temperature dependence is observed for CeBi, where the peak intensity of the reflectivity signal increases as the temperature decreases. In stark contrast, the peak intensity of the reflectivity signal in LaBi shows minimal changes with varying temperature. A fundamental distinction also lies in the non-oscillating signal's polarity: LaBi's non-oscillating component is consistently negative [Fig. 1(c)], whereas CeBi's remains perpetually positive [Fig. 1(a)]. This fundamental difference in $\Delta R(t)/R$ polarity and the characteristics of their relaxation pathways emphasise the divergent ultrafast carrier dynamics between CeBi and LaBi. These behaviors are further corroborated by the two-dimensional (2D) pseudocolor maps presented in Figs. 1(b) and 1(d).

To precisely isolate the coherent oscillatory component in CeBi, the non-oscillatory decay was meticulously fitted using a double-exponential function:

$$\frac{\Delta R(t)}{R} = H(\sigma, t)\left[\sum_{i=1}^{2} A_i \exp\left(-\frac{t}{\tau_i}\right) + C\right] \qquad (1)$$

Here, $H(\sigma, t)$ represents the Heaviside step function with an effective rise time $\sigma$. $A_i$ and $\tau_i$ denote the amplitude and relaxation time of the $i$-th decay process, respectively, while $C$ is a constant offset. As illustrated by the green solid curves in Fig. 1(a), Eq. (1) accurately describes CeBi's non-oscillatory response. This transient is characterized by an initial fast decay ($\tau_1 \sim 0.3$ ps) followed by a slower recovery ($\tau_2 \sim 1.6$ ps). In contrast, the non-oscillatory component of LaBi [Fig. 1(c)] exhibits a simpler, single-exponential decay ($\tau \sim 2.8$ ps). Given the semimetallic properties of both CeBi and LaBi, we posit that the observed carrier relaxation dynamics are primarily governed by electron-phonon and phonon-phonon scattering processes [34]. It is important to acknowledge that the presence of strong oscillation signals can introduce significant errors into the fitted parameters, which consequently complicates the accurate determination of how magnetic order transitions affect the measured relaxation times and amplitudes.

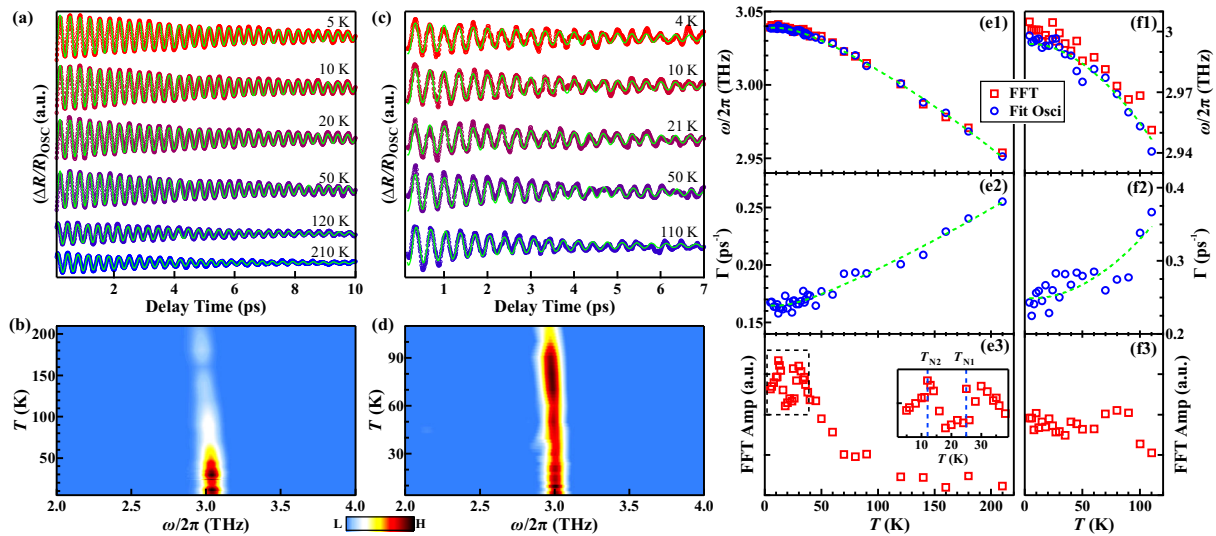

FIG. 2. (a, c) Extracted coherent oscillatory component of $\Delta R(t)/R$ for CeBi and LaBi, respectively, at various temperatures. Green solid curves represent fits using Eq. (2). (b, d) 2D map of FFT spectra as a function of temperature and frequency for CeBi and LaBi, respectively. (e, f) Temperature dependence of the extracted phonon frequency ($\omega/2\pi$), damping rate ($\Gamma$), and FFT amplitude for CeBi (e1-e3) and LaBi (f1-f3), respectively. Green dashed lines in (e1), (e2), (f1), and (f2) represent fits to the data using the anharmonic phonon model. The inset in (e3) highlights the FFT amplitude anomalies near the AFM transition temperatures $T_{N1}$ and $T_{N2}$.

The $\Delta R(t)/R$ signal oscillations were meticulously analyzed to identify collective excitations in both CeBi and LaBi. Figures 2(a) and 2(c) present the extracted time-domain oscillations for CeBi and LaBi, respectively, at various temperatures. Coherent oscillations are unequivocally observed across the entire measured temperature range in both compounds. Figures 2(b) and 2(d) further provide 2D maps of the Fast Fourier Transform (FFT) spectra, illustrating their dependence on temperature and frequency for CeBi and LaBi, respectively. For CeBi, a prominent coherent phonon with a frequency of approximately 3.04 THz (equivalent to 12.6 meV or 100.4 cm⁻¹) is identified at 5 K. This observation aligns well with theoretical predictions and previous reports, which identify this as the longitudinal optical (LO) coherent phonon, specifically the $A_{1g}$ mode, primarily stemming from the coherent vibrations of bismuth (Bi) atoms [35, 36]. Similarly, LaBi exhibits a coherent phonon at approximately 3.00 THz (12.4 meV or 100.1 cm⁻¹) at 4 K, which we also attribute to the $A_{1g}$ LO coherent phonon [35]. This conclusion contrasts with the transverse optical (TO) coherent phonon ($E_g$ mode) previously proposed by Yu et al. [37]. The significant temperature dependence observed in the frequencies of both $A_{1g}$ modes in CeBi and LaBi, characterized by a clear blueshift as temperature decreases [Figs. 2(b) and 2(d)], indicates the strong electron-phonon or phonon-phonon coupling present in these systems. Additionally, the FFT amplitude of these coherent phonons also exhibits a notable dependence on temperature, providing further insight into their underlying dynamics.

To quantitatively elucidate the temperature evolution of these coherent phonons, their frequencies ($\omega$) and damping rates ($\Gamma$) were precisely extracted by fitting the damped oscillations, shown as green solid lines in Figs. 2(a) and 2(c), using the expression:

$$(\Delta R(t)/R)_{Osc} = Ae^{-\Gamma t}\sin(\omega t + \phi), \quad (2)$$

Here, $A$, $\Gamma$, $\omega$, and $\phi$ represent the phonon amplitude, damping rate, frequency, and initial phase, respectively. The extracted frequency and damping rate for CeBi as a function of temperature are presented in Figs. 2(e1) and 2(e2), respectively, while the corresponding data for LaBi are shown in Figs. 2(f1) and 2(f2). A strong agreement is observed between the $A_{1g}$ mode frequencies estimated via FFT and those obtained from the damped oscillation fits for both CeBi and LaBi, as plotted in Figs. 2(e1) and 2(f1).

For CeBi, the $A_{1g}$ mode frequency continuously decreases from approximately 3.04 THz at 5 K to about 2.95 THz at 210 K with increasing temperature. Concurrently, its damping rate steadily increases with rising temperature. Similarly, for LaBi, the $A_{1g}$ mode frequency also exhibits a decrease with increasing temperature, shifting from around 3.00 THz at 4 K to approximately 2.94 THz at 110 K, with a notably faster rate of decrease compared to CeBi. Its damping rate similarly displays a consistent increasing trend. These observed behaviors are characteristic of optical phonon anharmonic ef-fects [38, 39], where increased thermal energy leads to enhanced phonon-phonon scattering and subsequent frequency softening and linewidth broadening. Such trends are consistent with observations in other semimetals like Bi [40], Sb [41] and PtTe$_2$ [42]. The temperature dependence of the $A_{1g}$ mode frequency and damping rate in both CeBi and LaBi can be accurately modeled by incorporating optical phonon anharmonic decay pathways, as evidenced by the excellent fits shown as green dashed lines in Figs. 2(e1), 2(e2), 2(f1) and 2(f2) across the entire measured temperature range. These observations strongly suggest that the relaxation of the $A_{1g}$ coherent phonon is predominantly governed by anharmonic decay into lower-energy acoustic phonons.

Figures 2(e3) and 2(f3) present the FFT amplitudes of the $A_{1g}$ modes for CeBi and LaBi, respectively. For LaBi, the FFT amplitude initially increases as temperature decreases, then plateaus, remaining nearly constant with further temperature reduction. In contrast, CeBi's FFT amplitude exhibits minimal change above approximately 120 K but undergoes a dramatic increase upon further cooling. This enhancement is particularly pronounced as the system approaches its magnetic ordering temperatures. As shown in the inset of Fig. 2(e3), the FFT amplitude displays clear anomalies near the two AFM transition temperatures, $T_{N1}$ and $T_{N2}$. Specifically, as the temperature drops below $T_{N1}$, the FFT amplitude anomalously decreases, subsequently rising again to reach a peak around $T_{N2}$, followed by another decrease for temperatures below $T_{N2}$. This distinct non-monotonic behavior strongly suggests robust magneto-elastic coupling between the $A_{1g}$ phonon mode and the magnetic order in CeBi. Such strong magneto-phononic coupling, where lattice dynamics are intimately intertwined with magnetic degrees of freedom, is a hallmark of many AFM compounds [8, 9, 17, 43–45]. This observation further demonstrates the exquisite sensitivity of coherent phonon dynamics as a probe for magnetic phase transitions in heavy fermion systems [15–17]. From a heavy-fermion perspective, this coupling is particularly insightful: the itinerant $f$-electrons, responsible for the magnetic ordering, can directly influence the lattice through exchange interactions, leading to observable phonon anomalies. Furthermore, in the context of topological materials, such strong magneto-phononic coupling could potentially induce or modify topological properties—for instance, by opening gaps or altering band inversions—if the phonon modes themselves carry topological character or interact with topologically protected electronic states. In this scenario, ultrafast spectroscopy offers a unique window into the dynamic interplay between these intertwined orders.

To further elucidate the nature of these coherent phonons and their intricate interaction with the electronic and magnetic degrees of freedom, we investigated their dependence on pump fluence. By systematically varying the photoexcitation intensity, our objective was to identify potential photoinduced phase transitions and assess the extent to which these transitions can be driven by ultrafast optical pulses. Figures 3(a1)-3(a3) present the pump fluence dependence of the frequency, FFT amplitude, and damping rate of the coherent $A_{1g}$ phonon mode in CeBi at 4 K (red markers) and 50 K (blue markers), respectively. The corresponding measurements for LaBi are displayed in Figs. 3(b1)-3(b3).

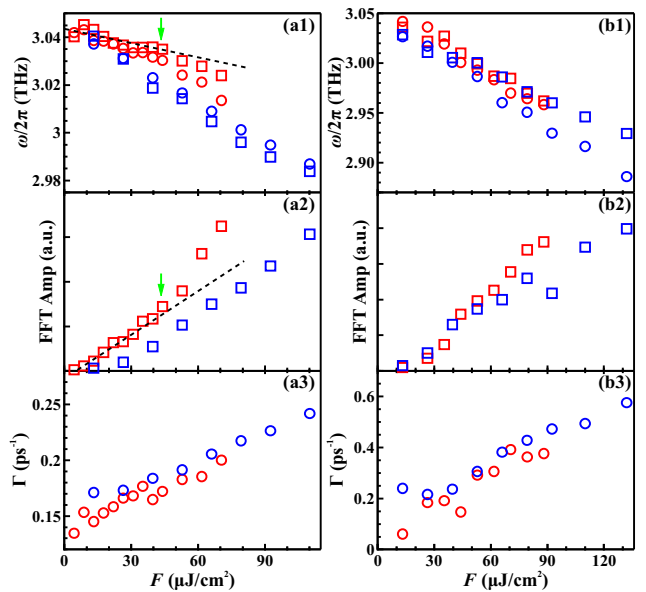

FIG. 3. (a1-a3) Extracted phonon frequency ($\omega/2\pi$), FFT amplitude, and damping rate ($\Gamma$), respectively, as a function of pump fluence ($F$) for CeBi at 4 K (red markers) and 50 K (blue markers). The black dashed lines serve as a guide for the eyes, and green arrows in (a1) and (a2) indicate the critical fluence threshold $F_C$. (b1-b3) Extracted phonon frequency ($\omega/2\pi$), FFT amplitude, and damping rate ($\Gamma$), respectively, as a function of pump fluence ($F$) for LaBi at 4 K (red markers) and 50 K (blue markers).

For CeBi at 4 K, a temperature deep within its AFM state, the $A_{1g}$ phonon frequency begins to decrease, and its FFT amplitude linearly increases with rising pump fluence. Intriguingly, above a critical fluence threshold, $F_C \approx 44 \ \mu\text{J/cm}^2$ (indicated by the green arrows in Figs. 3(a1) and 3(a2)), the rate of change for both parameters shifts dramatically: the phonon frequency decreases more rapidly, and the FFT intensity increases more sharply. In contrast, the phonon damping rate $\Gamma$ for CeBi at 4 K exhibits a monotonic linear increase with increasing pump fluence, suggesting strong carrier-phonon coupling, but displays no significant anomaly or change in slope at $F_C$. Conversely, for LaBi at 4 K, no anomalous behavior is observed in its phonon frequency, FFT amplitude, or damping rate across the entire measured fluence range. Moreover, the rate at which the phonon frequency decreases with increasing fluence in LaBi is significantly higher—more than three times as large—compared to CeBi.

Moving to 50 K, where CeBi is in its paramagnetic state, the frequency of its $A_{1g}$ phonon continuously decreases, and its FFT amplitude monotonically increases with increasing pump fluence. Importantly, no anomalies are observed over the entire measured fluence range at this temperature, indicating the absence of a photoinduced transition. CeBi's phonon damping rate at 50 K also increases linearly with fluence, similar to the 4 K case, and shows no discernible anomaly. However, the rate at which the phonon frequency decreases with increasing fluence in the paramagnetic state is significantly faster than that in the AFM state. Similarly, for LaBi at 50 K, consistent with its behavior at 4 K, no abnormal changes are observed in the phonon frequency, FFT intensity, or damping rate within the measured fluence range. Furthermore, the rate at which the phonon frequency decreases with increasing fluence in LaBi at 50 K remains significantly higher than that in CeBi—more than twice as large.

What accounts for the striking fluence dependence of such a phonon anomaly in CeBi within its AFM state? To elucidate the origin of this fluence threshold $F_C$, we considered two primary possibilities. First, intense pumping could induce a significant heating effect, particularly at low temperatures, potentially increasing the temperature of the illuminated sample region and consequently disrupting the AFM order. However, based on a simple steady-state heat diffusion model [46], and incorporating the material's optical constants [47], even considering a worst-case scenario (base temperature of 4 K and excitation density of ∼70.5 $\mu$J/cm$^2$), the calculated average laser-induced temperature rise in CeBi is less than 0.1 K. Thus, a simple laser heating effect can be definitively ruled out as the primary cause for the observed anomaly. Instead, the most compelling explanation points towards a photoinduced, non-thermal phase transition, potentially involving changes in the magnetic order. Given previous experimental work demonstrating the remarkable tunability of CeBi's magnetic order by external stimuli such as pressure [48, 49] and magnetic fields [50, 51], we attribute the observed phonon anomalies at $F_C$ to a rapid, non-thermal phase transition driven by intense photoexcitation—specifically, the quenching of the AFM order [30, 43, 44, 52]. This scenario aligns well with the unique characteristics of heavy fermion systems, where the delicate balance between localized magnetic moments and itinerant electrons can be readily perturbed by ultrafast optical pulses, leading to highly non-equilibrium states [17]. The coherent phonon, acting as a sensitive probe, immediately reflects these electronic and magnetic rearrangements. Further studies employing techniques like time-resolved X-ray diffraction or magneto-optical Kerr effect measurements, ideally with varying pump polarization or photon energy, would be invaluable to comprehensively elucidate the exact nature of these photoinduced magnetic phase transitions and their implications for the electronic band structure, potentially revealing photoinduced topological states or changes in the Fermi surface geometry.

In conclusion, our ultrafast optical spectroscopy investigation of CeBi and LaBi has provided significant insights into the intricate interplay among lattice dynamics, electronic correlations, and magnetic order within these semimetallic systems. We observed distinct coherent $A_{1g}$ phonon behaviors in CeBi compared to its non-$4f$ isostructural compound LaBi, particularly highlighting the strong magneto-phononic coupling evident in CeBi's phonon anomalies near its AFM transition temperatures. Furthermore, the discovery of a photoinduced, non-thermal quenching of AFM order in CeBi at a critical pump fluence highlights the potential for ultra-fast light pulses to dynamically manipulate magnetic states in correlated electron materials. These findings not only deepen our understanding of fundamental quasiparticle interactions in heavy fermion compounds but also establish coherent phonons as a powerful, sensitive probe for real-time monitoring of photoinduced phase transitions. Our work thus paves the way for future studies on the non-equilibrium control of complex quantum phases, including those with topological implications, in rare-earth monopnictides.

### ACKNOWLEDGEMENTS

This work was supported by the National Key Research and Development Program of China (Grant Nos. 2022YFA1604204 and 2024YFA140840), the National Natural Science Foundation of China (Grant No. 12074436), the Natural Science Foundation of Jiangsu Province (Grant No. BK20210276), the Open Project of Beijing National Laboratory for Condensed Matter Physics (Grant Nos. 2024BNL-CMPKF001 and 2023BNLCMPKF002), the Science and Technology Innovation Program of Hunan Province (Grant No. 2022RC3068), and the Synergetic Extreme Condition User Facility (SECUF, https://cstr.cn/31123.02.SECUF).

---

* jqmeng@csu.edu.cn

[1] J. Bardeen, L. N. Cooper, and J. R. Schrieffer, Phys. Rev. **108**, 1175 (1957).
[2] S. Yotsuhashi, M. Kojima, H. Kusunose, and K. Miyake, J. Phys. Soc. Jpn. **74**, 49 (2005).
[3] T. Hotta, Phys. Rev. Lett. **96**, 197201 (2006).
[4] M. Jaime, and M. B. Salamon, Rev. Mod. Phys. **73**, 583 (2001).
[5] S. Cheong, and M. Mostovoy, Nat. Mater. **6**, 13 (2007).
[6] C. Broholm, R. J. Cava, S. A. Kivelson, D. G. Nocera, M. R. Norman, and T. Senthil, Science **367**, eaay0668 (2020).
[7] S. Mi, J. Guo, G. Hu, G. Wang, S. Li, Z. Gong, S. Jin, R. Xu, F. Pang, W. Ji, W. Yu, X. Wang, X. Wang, H. Yang, and Z. Cheng, Nano Lett. **24**, 13094 (2024).
[8] H. Padmanabhan, M. Poore, P. K. Kim, N. Z. Koocher, V. A. Stoica, D. Puggioni, H. Wang, X. Shen, A. H. Reid, M. Gu, M. Wetherington, S. H. Lee, R. D. Schaller, Z. Mao, A. M. Lindenberg, X. Wang, J. M. Rondinelli, R. D. Averitt, V. Gopalan, Nat. Commun. **13**, 1929 (2022).
[9] T. T. Mai, K. F. Garrity, A. McCreary, J. Argo, J. R. Simpson, V. Doan-Nguyen, R. V. Aguilar, and A. R. H. Walker, Science **7**, eabj3106 (2021).
[10] T. Dong, S.-J. Zhang, and N.-L. Wang, Adv. Mater. **35**, 2110068 (2023).
[11] M. Porer, U. Leierseder, J. M. Ménard, H. Dachraoui, L. Mouchliadis, I. E. Perakis, U. Heinzmann, J. Demsar, K. Rossnagel, and R. Huber, Nat. Mater. **13**, 857 (2014).
[12] Y. Wang, B. Chen, Z. Liu, X. Guo, S. You, Z. Wang, H. Xie, T. Niu, J. Meng, and H. Huang, Phys. Rev. B **108**, 045118 (2023).
[13] M. Le Tacon, A. Bosak, S. M. Souliou, G. Dellea, T. Loew, R. Heid, K. Bohnen, G. Ghiringhelli, M. Krisch, and B. Keimer, Nat. Phys. **10**, 52 (2014).
[14] Q. Y. Wu, C. Zhang, B. Z. Li, H. Liu, J. J. Song, B. Chen, H. Y. Liu, Y. X. Duan, J. He, J. Liu, G. H. Cao, and J. Q. Meng, Phys.

Rev. B **111**, L081110 (2025).

[15] K. S. Burch, E. E. M. Chia, D. Talbayev, B. C. Sales, D. Mandrus, A. J. Taylor, and R. D. Averitt, Phys. Rev. Lett. **100**, 026409 (2008).

[16] Y. P. Liu, Y. J. Zhang, J. J. Dong, H. Lee, Z. X. Wei, W. L. Zhang, C. Y. Chen, H. Q. Yuan, Y. F. Yang, and J. Qi, Phys. Rev. Lett. **124**, 057404 (2020).

[17] Y. Z. Zhao, Q. Y. Wu, C. Zhang, B. Chen, W. Xia, J. J. Song, Y. H. Yuan, H. Liu, F. Y. Wu, X. Q. Ye, H. Zhang, H. Huang, H. Y. Liu, Y. X. Duan, Y. F. Guo, J. He, and J. Q. Meng, Phys. Rev. B **108**, 075115 (2023).

[18] F. F. Tafti, Q. D. Gibson, S. K. Kushwaha, N. Haldolaarachchige, and R. J. Cava, Nat. Phys. **12**, 272 (2016).

[19] L. Ye, T. Suzuki, C. R. Wicker, and J. G. Checkelsky, Phys. Rev. B **97**, 081108 (2018).

[20] F. F. Tafti, M. S. Torikachvili, R. L. Stillwell, B. Baer, E. Stavrou, S. T. Weir, Y. K. Vohra, H. Y. Yang, E. F. McDonnell, S. K. Kushwaha, Q. D. Gibson, R. J. Cava, and J. R. Jeffries, Phys. Rev. B **95**, 014507 (2017).

[21] R. Lou, B.-B. Fu, Q. N. Xu, P.-J. Guo, L.-Y. Kong, L.-K. Zeng, J.-Z. Ma, P. Richard, C. Fang, Y.-B. Huang, S.-S. Sun, Q. Wang, L. Wang, Y.-G. Shi, H. C. Lei, K. Liu, H. M. Weng, T. Qian, H. Ding, and S.-C. Wang, Phys. Rev. B **95**, 115140 (2017).

[22] C. Y. Guo, C. Cao, M. Smidman, F. Wu, Y. J. Zhang, F. Steglich, F. C. Zhang, and H. Q. Yuan, npj Quant. Mater. **2**, 39 (2017).

[23] P. Li, Z. Z. Wu, F. Wu, C. Y. Guo, Y. Liu, H. J. Liu, Z. Sun, M. Shi, F. Rodolakis, J. L. McChesney, C. Cao, H. Q. Yuan, F. Steglich, and Y. Liu, Phys. Rev. B **100**, 155110 (2019).

[24] T. Kasuya, Y. Haga, T. Suzuki, Y. Kaneta, and O. Sakai, J. Phys. Soc. Jpn. **61**, 3447 (1992).

[25] Y. Y. Lyu, F. Han, Z. Li Xiao, J. Xu, Y. L. Wang, H. B. Wang, J. K. Bao, D. Y. Chung, M. Li, I. Martin, U. Welp, M. G. Kanatzidis, and W. K. Kwok, Phys. Rev. B **100**, 180407 (2019).

[26] B. J. Feng, J. Cao, M. Yang, Y. Feng, S. L. Wu, B. T. Fu, M. Arita, K. Miyamoto, S. L. He, K. Shimada, Y. G. Shi, T. Okuda, and Y. G. Yao, Phys. Rev. B **97**, 155153 (2018).

[27] J. Jiang, N. B. M. Schr Oter, S. C. Wu, N. Kumar, C. Shekhar, H. Peng, X. Xu, C. Chen, H. F. Yang, C. C. Hwang, S. K. Mo, C. Felser, B. H. Yan, Z. K. Liu, L. X. Yang, Y. L. Chen, Phys. Rev. Materials **2**, 024201 (2018).

[28] A. de la Torre, D. M. Kennes, M. Claassen, S. Gerber, J. W. McIver, and A. M. Sentef, Rev. Mod. Phys. **93**, 041002 (2021).

[29] S. Z. Zhao, H. Y. Song, L. L. Hu, T. Xie, C. Liu, H. Q. Luo, C. Y. Jiang, X. Zhang, X. C. Nie, J. Q. Meng, Y. X. Duan, S. B. Liu, H. Y. Xie, and H. Y. Liu, Phys. Rev. B **102**, 144519 (2020).

[30] H. Liu, C. Zhang, Q.-Y. Wu, Y. Jin, Z. Zhu, J.-J. Song, S.-T. Cui, Z. Sun, H. Wang, B. Chen, J. He, H.-Y. Liu, Y.-X. Duan, P. M. Oppeneer, and J.-Q. Meng, Phys. Rev. B **111**, L121113 (2025).

[31] C. Zhang, Q. Y. Wu, W. S. Hong, H. Liu, S. X. Zhu, J. J. Song, Y. Z. Zhao, F. Y. Wu, Z. T. Liu, S. Y. Liu, Y. H. Yuan, H. Huang, J. He, S. L. Li, H. Y. Liu, Y. X. Duan, H. Q. Luo, and J. Q.

Meng, Sci. China-Phys. Mech. Astron. **65**, 237411 (2022).

[32] B. L. Tan, C. Zhang, Q. Y. Wu, G. H. Dong, H. Liu, B. Chen, J. J. Song, X. Y. Tian, Y. Zhou, H. Y. Liu, Y. X. Duan, Y. G. Shi, and J. Q. Meng, Front. Phys. **20**, 044208 (2025).

[33] S. C. Huan, X. B. Shi, L. Han, H. Su, X. Wang, Z. Q. Zou, N. Yu, W. W. Zhao, L. M. Chen, Y. F. Guo, J. Alloys Compd. **875**, 159993 (2021).

[34] Y. M. Dai, J. Bowlan, H. Li, H. Miao,S. F. Wu,W. D. Kong, P. Richard,Y. G. Shi, S. A. Trugman, J. -X. Zhu, H. Ding, A. J. Taylor, D. A. Yarotski, and R. P. Prasankumar, Phys. Rev. B **92**, 161104(R) (2015).

[35] V. H. Mankad, S. K. Gupta, I. Lukacevic, and P. K. Jha, Comp. Mater. Sci. **65**, 536 (2012).

[36] E. Mariani, Lattice dynamics and phonon anomalies in Cerium compounds: The particular case of CeBi and CeSb, Master Thesis, 2017.

[37] B. H. Yu, Z. Y. Tian, F. Sun, D. C. Peets, X. D. Bai, D. L. Feng, and J. Zhao, Opt. Express **28**, 15855 (2020).

[38] M. Balkanski, R. F. Wallis, and E. Haro, Phys. Rev. B **28**, 1928 (1983).

[39] J. Menéndez, and M. Cardona, Phys. Rev. B **29**, 2051 (1984).

[40] M. Hase, M. Kitajima, S. Nakashima, and K. Mizoguchi, Phys. Rev. Lett. **88**, 067401 (2002).

[41] M. Hase, K. Ushida, and M. Kitajima, J. Phys. Soc. Jpn. **84**, 024708 (2015).

[42] Z. Li, Y. Chen, A. Song, J. Zhang, R. Zhang, Z. Zhang, and X. Wang, Light Sci. Appl. **13**, 181 (2024).

[43] D. Afanasiev, J. R. Hortensius, B. A. Ivanov, A. Sasani, E. Bousquet, Y. M. Blanter, R. V. Mikhaylovskiy, A. V. Kimel, and A. D. Caviglia, Nat. Mater. **20**, 607 (2021).

[44] H. Liu, Q.-Y. Wu, C. Zhang, J. Pang, B. Chen, J.-J. Song, Y.-X. Duan, Y.-H. Yuan, H.-Y. Liu, C.-C. Shu, Y.-F. Xu, Y.-G. Shi, and J.-Q. Meng, Phys. Rev. B **110**, 195104 (2024).

[45] L. Cheng, T. Xiang, and J. Qi, Phys. Rev. Research **6**, 023073 (2024).

[46] J. Demsar, J. L. Sarrao, and A. J. Taylor, J. Phys.: Condens. Matter **18**, R281 (2006).

[47] Y. S. Kwon, H. Kitazawa, N. Sato, H. Abe, T. Nanba, M. Ikezawa, K. Takegahara, O. Sakai, T. Suzuki, and T. Kasuya, Jpn. J. Appl. Phys. **26**, 539 (1987).

[48] H. Bartholin, P. Burlet, S. Quezel, J. Rossat-Mignod, O. Vogt, Le Journal de Physique Colloques **40**, C5-130 (1979).

[49] P. H. Hor, R. L. Meng, S. Yomo, C. W. Chu, E. Bucher, P. H. Schmidt, Physica **139-140**, 378 (1986).

[50] B. Kuthanazhi, N. H. Jo, L. Xiang, S. L. Bud'Ko, and P. C. Canfield, Philos. Mag. **102**, 542 (2021).

[51] J. Rossat-Mignod, P. Burlet, S. Quezel, J. M. Effantin, D. Delacote, H. Bartholin, O. Vogt, and D. Ravot, J. Magn. Magn. Mater. **31-34**, 398 (1983).

[52] A. Kirilyuk, A. V. Kimel, and T. Rasing, Rev. Mod. Phys. **82**, 2731 (2010).